\documentstyle[prb,preprint,aps]{revtex}
\draft
\input{psfig}
\psfigurepath{FIG/}

\def\AmS{{\protect\the\textfont2
        A\kern-.1667em\lower.5ex\hbox{M}\kern-.125emS}}

\makeatletter
\def\thepage{1-\@arabic\c@page}
\def\@pnumwidth{2em}
\makeatother
\begin{document}

\title{Transmission Through Carbon Nanotubes With Polyhedral Caps}
\author{M. P. Anantram~\cite{byline1} and T. R. Govindan\\}
\address{ NASA Ames Research Center, Mail Stop T27A-1,
           Moffett Field, CA, USA 94035-1000  }

\maketitle

\begin{abstract}
We study electron transport between capped carbon nanotubes and a 
substrate, and relate the transmission probability to the local density
of states in the cap. Our results show that the transmission 
probability mimics the behavior of the density of states at all 
energies except those that correspond to localized states in the cap.
Close proximity of a substrate causes hybridization of the localized 
state. As a result, new transmission paths open from the substrate to
nanotube continuum states via the localized states in the cap. 
Interference between various transmission paths gives rise to {\it 
antiresonances} in the transmission probability, with the minimum 
transmission equal to zero at energies of the localized states. Defects
in the nanotube that are placed close to the cap cause {\it resonances}
in the transmission probability, instead of antiresonances, near the 
localized energy levels. Depending on the spatial position of defects,
these resonant states are capable of carrying a large current. These 
results are relevant to carbon nanotube based studies of molecular 
electronics and probe tip applications.

\vspace{0.2in}

Physical Review B, vol. 61, p. 5020 (2000)

\end{abstract}
\pacs{}

\section{Introduction}

Characteristic features of electron flow through nanotubes are relevant
to both molecular electronics and experiments using nanotube tips as a 
probe. 
In these applications, the nanotube tips can be capped or open along 
with appropriate functionalization if desired. In preliminary studies,
Dai et. al~\cite{Dai96} and Wong et. al~\cite{Wong98b} used nanotube 
tips for high resolution imaging, and Wong et. al~\cite{Wong98a} have
extended their study to include functionalized 
tips.~\cite{Wong98a,comment1}
In view of such studies, which can in principle be extended to STM 
measurements, it is important to understand the nature of electron
flow from nanotube tips. 
The large number of possible topological arrangements of carbon atoms
at the tip of a nanotube and the possibility of functionalizing tips
makes the study of electron transport an interesting and necessary
one. Further, nanotube tips have recently been observed by various
authors,~\cite{Carroll97,Kim99,footnote_cap} and methods of 
constructing caps have been suggested.~\cite{Fujita92}
To the best of our knowledge, electron transport through capped 
nanotubes have not been studied, although the local density
of states (LDOS) have been studied.~\cite{Carroll97,Kim99,Tamura95}
In this paper, we study electron flow from a substrate to a nanotube 
tip, much like in an STM experiment where the nanotube is the tip.
In this particular study, we restrict the topology of the tip to 
that of a polyhedral cap. The LDOS at the cap has resonances 
corresponding to quasi-localized states as observed experimentally in
Refs.  \onlinecite{Carroll97} and \onlinecite{Kim99}.
In particular, Tamura et. al~\cite{Tamura95} have theoretically shown
the existence of purely localized states in nanotubes with polyhedral
caps. The effect of localized states on current flow, and the 
relationship between local density of states and current flow through
various cap atoms are of particular interest.

Consider a nanotube interacting with a substrate as shown in Fig. 1(a).
The wave functions of the cap and substrate overlap due to their 
physical proximity.~\cite{footnote1}
This overlap provides a physical mechanism for the hybridization of
localized and continuum states, which causes the localized states 
(discrete energy levels) to become quasi-localized.
The effect of localized states on the transmission probability in
general {\it depends} on the nature of interaction between the
localized and continuum states. Examples from the literature, which
illustrate this point are a quantum well with quasi-localized levels
(double barrier resonant tunneling diode) and a quantum wire
with a stub containing quasi-localized levels.~\cite{Shao94}
Consider a quantum well with a single localized level. Let this level
be coupled to continuum states both above and below as shown in Fig. 
1(b). An electron incident from the continuum states on the top can
be transmitted to the continuum states below only via the localized
state. It is well known that the transmission probability through such
a structure exhibits a resonance that corresponds to the localized
state. In contrast to this example, a localized state can also interact
with a continuum as shown in Fig. 1(c). The primary difference of
electron transmission paths in this case when compared to the double 
barrier structure in Fig. 1(b) is that an electron incident in the
substrate can be transmitted to the continuum states of the nanotube 
via paths that do not use the localized state (in a perturbative 
sense), in addition to paths that use the localized state.
Similar transmission paths exist in the context of scattering of light
from molecules,~\cite{Fano61} electron transport through quantum wires
with stubs~\cite{Shao94} and tunneling through a
heterostructure barrier.~\cite{Bagwell92} In these cases, localized 
states play an important role in determining the transmission 
probability around the localized energy level. In particular, the 
transmission probability exhibits an antiresonance due to the localized
energy level.

An isolated nanotube with a polyhedral cap has localized states in the
cap that decay into the nanotube.~\cite{Tamura95}
Electrons can be transmitted from the substrate to the nanotube by 
paths that both do and do not use the localized level. We show that
the transmission probability in this case has an
{\it antiresonance} corresponding to the energy of the localized level.
This picture changes drastically if there are defects in the nanotube.
Defects in the nanotube as shown pictorially in Fig. 1(d) open new
transmission paths, which cause {\it resonances} in the
transmission probability close to the localized energy levels.

We focus on the truly metallic armchair tubes, which show promise as
quantum wires and for CNT based probes involving a tunnel current.
The 5-fold symmetric polyhedral cap with one pentagon at the cap center
and five pentagons placed symmetrically along the edge of a (10,10)
nanotube is considered [Fig. 2]. The outline of the paper is as
follows.  In section II, we describe the method used to calculate
the transmission probability and density of states. 
In section III, we study the relationship between the LDOS and
electron transmission probability by addressing the following issues:
(i) the relationship between LDOS and transmission probability through
cap atoms,
(ii) the effect of the localized energy levels in the cap (section 
IIIA),
(iii) a simple one dimensional model to understand the essential 
feature of sections IIIA and IIIC  (section IIIB), and
(iv) the effect of defects on tunnel current/transmission probability
(section IIIC).
Conclusions of this study are summarized in section IV.

\section{Method}

In this section we outline the formalism used and also discuss the 
assumptions made in our study. 
The combination of CNT and substrate can be conceptually divided into
three regions: substrate ({\bf S}), section of CNT including the cap 
({\bf D}) and a semi-infinite CNT region ({\bf L}) as shown in Fig.
1(a). The advantage of this procedure is that the influence of the
semi-infinite region {\bf L} and the substrate can be included exactly
as a self-energy to the $[E-H]$ matrix with dimension equal to the
number of atoms in {\bf D} ($H$ is the Hamiltonian of
{\bf D}).~\cite{Caroli71}
This procedure is not sensitive to the exact location of the interface
between {\bf L} and {\bf D} when charge self-consistency is neglected.
We typically take region $D$ to consist of five to a hundred
unit cells of an armchair tube, and the results are not sensitive to 
the exact number as long as the retarded Green's function of region
{\bf L} (see $g^r_L$ below) is calculated accurately.
The transmission and LDOS are calculated using the formalism in 
reference \onlinecite{Caroli71} and adapted for nanotubes in
reference \onlinecite{Anantram98}.
The retarded Green's function $G^r$ is obtained by solving:
\begin{eqnarray}
\left[ E - H - \Sigma^r_L(E) - \Sigma^r_S(E) \right] G^r(E) =
                                           I \mbox{,} \label{eq:xxx}
\end{eqnarray}
where $H$ is the sub-Hamiltonian of region {\bf D}. 
$\Sigma^r_L (E) = V_{D L} g^r_L (E) V_{L D}$ is
the self energy due to the semi-infinite region {\bf L} that is
folded into the Hamiltonian of region {\bf D}.
$V_{DL}$ ($V_{LD}$) is the term in the full Hamiltonian representing 
the interaction between {\bf D} ({\bf L}) and {\bf L} ({\bf D}), and 
$g^r_L$ is the retarded Green's function of region {\bf L} that is
calculated by assuming {\bf L} to be isolated from {\bf D} and {\bf S}.
The only terms of $g^r_L$ that enter the calculation of
$\Sigma^r_L (E)$ corresponds to lattice sites in {\bf L} that are
connected to {\bf D}. That is, only the surface Green's function of 
{\bf L} is required.~\cite{Anantram98}
$\Sigma^r_S$ is the self energy due to the substrate. 
In this paper, we assume $\Sigma^r_S$ to be an energy independent
parameter that represents coupling between {\bf S} and only one atom 
in the cap.
This assumption is usually valid over small energy ranges that 
are away from sharp features in the substrate density of states.

The single particle LDOS at site $i$ [$N_i(E)$] and  transmission 
probability [$T(E)$] at energy $E$ are obtained by solving 
Eq. (\ref{eq:xxx}) for the diagonal element $G^r_{ii}$ and the 
corner off-diagonal sub-matrix of $G^r$ whose row and column indices
correspond to atoms in {\bf D} that couple {\bf L} and {\bf S}
respectively:
\begin{eqnarray}
N_i(E) &=& -\frac{1}{\pi} Im[ G^r_{ii} (E) ]  \\
  T(E) &=& Trace[\Gamma_{L} G^r \Gamma_{S} G^a] \mbox{ .} 
\label{eq:Transmission}
\end{eqnarray}
$\Gamma_L$ and $\Gamma_S$ are the coupling rates of {\bf D} to the 
semi-infinite nanotube {\bf L} and substrate {\bf S} respectively.
$\Gamma_L = 2 \pi V_{DL} \rho_L (E) V_{LD}$, where 
$\rho_L(E) = -\frac{1}{\pi} Im[g_L^r(E)]$ is the surface density of
states of {\bf L} ($Im$ extracts the imaginary part) and
$\Gamma_S = -2Im[\Sigma^r_S]$.

For tubes with defects, we consider the Stone-Wales model that creates
two pentagon-heptagon pairs in the hexagonal network (see dashed box in
Fig. 1).~\cite{Yakobson98,Dresselhaus_book} 
Finally, the numerical calculations use the single orbital real space
tight binding representation of the CNT 
Hamiltonian,\cite{Dresselhaus_book} 
\begin{eqnarray}
H = -b \sum_{i \neq j} c_i^\dagger c_j + c.c. \mbox{ ,}
\end{eqnarray}
where each carbon atom has a hopping parameter $b$ with its 
three near neighbors, and $c_i$ ($c_i^\dagger$) is the annihilation 
(creation) operator at atomic site $i$. 
The value of $b$ is chosen to be 3.1 eV.

\section{Results and Discussion}
\label{section:results}

The main issues addressed in this section are:
(i) the relationship between LDOS and transmission probability
through cap atoms in a defect free CNT,
(ii) the effect of localized energy levels on the transmission
probability (section \ref{subsect:results_antiresonance}),
(iii) a simple one dimensional model to understand the essential
features of sections \ref{subsect:results_antiresonance} and
\ref{subsect:results_defects}
(section \ref{subsect:results_simple_model}) and,
(iv) the effect of defects on tunnel current/transmission
probability (section \ref{subsect:results_defects}).
We only consider weak coupling between the nanotube and substrate
{\bf S}. The coupling strength ($\Sigma^r_S$) is defined to be
weak if it is smaller than the value between two near neighbor carbon
atoms along the length of the nanotube (diagonal terms of
$\Sigma^r_L$).

\subsection{Antiresonances in transmission probability}
\label{subsect:results_antiresonance}

We first address issues (i) and (ii) involving defect free caps by
studying the relationship between the LDOS at atom $i$ in the cap and
the transmission probability from the substrate to the semi-infinite
CNT via atom $i$. An isolated polyhedral cap has localized levels
(no broadening).~\cite{Tamura95} Fig. 3 shows the effect of coupling
of the cap to the substrate. In the calculations, one atom in the cap
couples to the substrate (as labeled in Fig. 3). 
Coupling of the cap to the substrate causes hybridization with
the substrate continuum states. As a result of hybridization, the 
localized states become quasi-localized, as represented by the 
broadened resonances in Fig. 3. In the energy range considered, there
are two localized states, one around 0.25 eV and the other around -1.5
eV. The value of $\Sigma^r_S = -i\:\Gamma_S/2 = - 12.5\:i\:$meV for all 
curves in Fig. 3.

The LDOS varies significantly with atomic location. The LDOS at
apex atom 1 is almost an order of magnitude smaller than the LDOS at 
atom 4 which is located at the cap edge. The LDOS of atoms 2 and 3 
which lie in between, and the DOS averaged over all cap atoms are also
shown for comparison.
The transmission probability versus energy for the cases corresponding
to Fig. 3 are shown in Fig. 4. The strength of coupling in Fig. 4 is 
the same as in Fig. 3.
The transmission probability follows the LDOS at most energies in that
the magnitude is proportional to the LDOS as is seen by comparing Figs.
3 and 4. There is a major difference at the resonant energy, where
{\it the LDOS peaks corresponds to transmission zeroes}.
The transmission antiresonances arise from hybridization of localized
and continuum states via coupling to the substrate as represented
pictorially in Fig. 1(c). States in the CNT cap comprise of localized
($\phi_l$) and continuum ($\phi_c$) states that are not coupled to each
other in an isolated cap.
Bringing the substrate in close proximity to the cap couples $\phi_l$
and $\phi_c$ to the substrate states ($\phi_s$).
As a result, electrons have many paths to be transmitted from $\phi_s$
to $\phi_c$: (i) directly from $\phi_s \rightarrow \phi_c$,
(ii)  $\phi_s \rightarrow \phi_l \rightarrow \phi_s \rightarrow \phi_c$
and (iii) higher order representations of (ii).
The interference between these paths gives rise to the transmission
zeroes at the resonant energies (inset of Fig. 4).
The numerical calculation in this subsection is complemented by a 
simplified analytical model in section 
\ref{subsect:results_simple_model} to demonstrate the physics more
clearly.

When the strength of coupling between the cap and substrate increases,
the antiresonances becomes more pronounced. That is, the minimum is 
still zero but the width increases with increase in coupling strength
$\Gamma_s$. To demonstrate this, atom 4 is assumed to make contact to
the substrate with coupling strengths as given by the legend of Fig. 5.
The real part of $\Sigma^r_S$ has been assumed to be zero in Figs. 3,
4 and 5. The primary effect of including a non zero real part of 
$\Sigma^r_S$ is to make the transmission versus energy more asymmetric.
The position of the antiresonance however remains unchanged.

\subsection{Simple one dimensional model}
\label{subsect:results_simple_model}

In this section, we present a simple one dimensional model of transport
from a tip that has a localized state to a substrate, much like in 
Fig. 1(a). The expression for transmission coefficient is obtained
analytically and aids in understanding the numerical results of 
sections \ref{subsect:results_antiresonance} and 
\ref{subsect:results_defects}, which consider a nanotube. We now define
the model. The continuum states of the tip are modeled as a one 
dimensional semi-infinite chain with on-site energy $\epsilon_c$ and 
hopping parameter $b_c$ (Fig. 6).
The energy spectrum of such a chain has a band width equal to $4b_c$ 
and there are no localized states in this band width.
The localized level at the tip is modeled as a state with energy 
$\epsilon_l$ that lies at the edge of the tip (Fig. 6) and within the
continuum of the tip states (energy-wise).
The localized state hybridizes with the continuum states of the tip
at the edge atom and this is modeled by a hopping parameter $t_{lc}$.
The localized state ($t_{lc}=0$) of an isolated tip becomes 
quasi-localized when $t_{lc} \neq 0$.
The substrate is also modeled as a one dimensional semi-infinite chain
with on-site potential $\epsilon_s$ and hopping parameter $b_s$
(Fig. 6). The substrate states hybridize with the continuum and 
localized states of the tip only at the tip-edge, and this is modeled
by hopping parameters $t_{cs}$ and $t_{ls}$ respectively.
The subscripts {\it l, c} and {\it s} represent localized, continuum 
and substrate states respectively.

Following the method described in section II,~\cite{Caroli71} the 
retarded Green's function of the continuum and localized states
at the edge of the tip is,
\begin{eqnarray}
G^r(E) = D^{-1} \cdot
\left(
\begin{array}{cc}
E - \epsilon_l - t_{ls}^2 g^r_s(E) & t_{lc} + t_{cs} t_{ls} g^r_s(E)  \\
t_{lc} + t_{cs} t_{ls} g^r_s(E)    & E - \epsilon_c - t_{cs}^2 g^r_s(E) - b_c^2
g^r_c(E)
\end{array}
\right)
\mbox{ ,} \label{eq:Gmodel}
\end{eqnarray}
where, D is the determinant of the matrix in the bracket of Eq.
(\ref{eq:Gmodel}). The (1,1) and
(2,2) components of $G^r(E)$ correspond to the continuum and localized
states respectively, and the off-diagonal term corresponds to the 
correlation between the continuum and localized states. The surface 
Green's function of the substrate and tip chains are given by,
$g^r_i =\frac{E-\epsilon_i-\sqrt{(E-\epsilon_i)^2-4b_i^2}}{2b_i^2}$,
where $i \in {c,s}$.

The transmission amplitude of an electron from the substrate to
the nanotube is,~\cite{Caroli71}
\begin{eqnarray}
t(E) = [2\pi\rho_c(E)]^{\frac{1}{2}} b_{c} \: G^r_{11}(E) \: t_{cs}
                                      [2\pi\rho_s(E)]^{\frac{1}{2}}
     + [2\pi\rho_c(E)]^{\frac{1}{2}} b_{c} \: G^r_{12}(E) \: t_{ls}
                                      [2\pi\rho_s(E)]^{\frac{1}{2}}
\mbox{ ,} \label{eq:smatrix}
\end{eqnarray}
where, $G^r_{11}$ and $G^r_{12}$ are the (1,1) and (1,2) elements
of the Green's function matrix (Eq. \ref{eq:Gmodel}), and $\rho_i(E)
=-\frac{1}{\pi}\mbox{Im}[g^r_i(E)]$, where $i\in s,c$ is the surface 
density of states.
The first term of Eq. (\ref{eq:smatrix}) represents the path where an
electron incident in the tip is transmitted to the substrate via the 
modified continuum states at the edge atom of the tip.
The modification of the continuum states are due to interaction with
the localized state and the substrate states.
The second term of Eq. (\ref{eq:smatrix}) represents the path where an
electron incident in the tip is transmitted to the substrate via the 
localized state in the edge atom.
$G_{12}^r(E)$ represents the correlation between the continuum and
localized levels at the edge atom of the tip when the tip is connected
to the substrate.
The transmission probability ($|t(E)|^2$) then consists of interference
between paths that do and do not use the localized state.

From Eqs. (\ref{eq:Gmodel}) and (\ref{eq:smatrix}), the following 
three observations can be made:
{\it (i)} The transmission coefficient is zero when,
\begin{eqnarray}
E = \epsilon_l - t_{lc}\:\frac{t_{ls}}{t_{cs}} \mbox{ .}
\label{eq:T_is_0}
\end{eqnarray}
As mentioned in the beginning of this section, the case of a localized
state corresponds to $t_{lc}=0$. Then, the transmission antiresonance
is at the energy of the localized state $\epsilon_l$.
An example is shown in Fig. 7(a), where the parameters are
$b_c=b_s=1\:$eV, $t_{ls}=t_{cs}=0.04\:$eV, $\epsilon_c=\epsilon_s=0\:$eV
and $\epsilon_l=0.245\:$eV.
{\it (ii)} From Eq. (\ref{eq:T_is_0}), it is clear that the
transmission probability has a zero in the presence of defects that
cause hybridization between the localized and continuum states
($t_{lc} \neq 0$) even when the tip does not make contact to the
substrate. The location of the transmission zero has however moved
away from the localized energy by $-t_{lc}\:\frac{t_{ls}}{t_{cs}}$.
{\it (iii)} A transmission resonance results when the localized level
hybridizes with the continuum states in the tip ($t_{lc}\neq0$). More
specifically, when the hopping parameters $t_{cs}$, $t_{ls}$ and 
$t_{lc}$ are much smaller than $b_c$ and $b_s$, the transmission
probability has a resonance at an energy close to the localized energy
level $\epsilon_l$ [Fig. 7(b)]. The precise location of the resonance
depends on the values of the hopping parameters. The width of the
resonance increases with increase in $t_{lc}$ [Fig. 7(b)], a feature
that is observed in the case of the nanotube also is discussed in the
following subsection. Except for $t_{lc}$, the parameters used in
Fig. 7(b) are the same as in Fig. 7(a).

\subsection{Influence of defects}
\label{subsect:results_defects}

We now consider the changes to the antiresonance picture discussed in
section \ref{subsect:results_antiresonance} due to defects in the
nanotube. The results for two different locations of a defect along the
length of the nanotube and atom 1 making contact to the substrate with
coupling strength $\Sigma_S^r = -i\,\Gamma_S/2 = - 25\,i\,$meV are 
shown in Fig. 8. While the defect considered here is the Stone-Wales 
defect (see dashed box in Fig. 1), we have also carried out similar 
calculations for defect models that involve a random change in the 
on-site potential and hopping parameter of carbon atoms located near 
the cap. The main conclusion of these calculations is that the results
of this section do not qualitatively depend on the exact defect model
as long as the defect hybridizes the localized and continuum states of
the nanotube.

Scattering due to the defect opens up more channels for hybridization
of the localized state. As a result, the LDOS of the cap will show
broadened resonances similar to those in Fig. 3.
The transmission probability changes significantly around the localized
energy levels in comparison to Fig. 4, as shown in Fig. 8. The sharp 
transmission antiresonances at the localized energy has disappeared and
instead the transmission probability has sharp resonances around the
energy of the localized state. This is because the  defect locally 
mediates mixing/hybridization of localized and continuum states. As a
result, the localized states are coupled to continuum states by two 
scattering centers: the defect in the nanotube and the interaction 
with the substrate.
This leads to additional transmission paths that are similar in spirit
to paths in a double barrier resonant tunneling structure in Fig. 1(b),
where the two scattering centers are the barriers. The simple model
discussed in section \ref{subsect:results_simple_model} also 
demonstrates this point (Fig.  7 and discussion of (iii) at the end of
section \ref{subsect:results_simple_model}).

The resonance width of the transmission probability is determined by 
two contributions.
The first contribution is the hybridization due to the substrate and 
the second contribution is the hybridization due to the defect.
The second contribution depends on $|<\phi_C|H_{defect}|\phi_L>|$, 
where $H_{defect}$ is Hamiltonian of the defect.
Fig. 8 shows the transmission probability for two different distances
of the defect from the cap ($L_D$).
$L_D$ equal to 7 and 15 are in units of the one dimensional unit cell
length of armchair tubes.
The main feature is that the {\it width becomes smaller as distance of
the defect from the cap increases}.
This trend can be understood from the fact that $|\phi_L|^2$ (or the 
density of states of the localized state) decays with distance away 
from the cap. As a result, the strength of hybridization
between continuum and localized states in the cap arising due to the
defect ($|<\phi_C|H_{defect}|\phi_L>|$) decreases as distance of the
defect from the cap apex increases. This corresponds to $t_{lc}$ 
becoming smaller in the model discussed in section 
\ref{subsect:results_simple_model} (see discussion of (iii) at the end
of section \ref{subsect:results_simple_model}).

\section{Conclusions}

We studied phase coherent transport through carbon nanotube tips in 
proximity to a substrate. An armchair tube with a polyhedral cap
has localized states that decay with distance away from the
cap. We find that these localized states play an 
important role in determining the features of the electron transmission
probability from the substrate to the nanotube. The transmission
probability corresponds directly to the LDOS at energies away from the
localized energy levels. Close to the localized energy level, while the
LDOS exhibits a resonance, the transmission probability exhibits an 
antiresonance.
Defects in the tube alter the antiresonance by providing additional 
defect-assisted channels for transport into the continuum states of the
CNT. As a result, the transmission probability has a resonance close to
the localized energy levels, instead of an antiresonance. These 
resonances are capable of carrying a large amount of current compared
to other energies, and so are relevant to experiments that measure the
tunnel current using carbon nanotube based tips. The current carrying
capacity of the resonance depends on two parameters: the hybridization
strengths of the localized state due to interaction with the substrate,
and defect assisted interaction with the continuum states
of the nanotube. Since the density of states of the localized levels
decay with distance into the nanotube, the hybridization strength due
to defect assisted scattering decreases. The current carrying capacity
of the resonances then decreases with increase in the distance of the
defect from the cap.

We would like to thank Liu Yang and Jie Han of NASA Ames Research 
Center for useful discussions, and Bryan Biegel for useful comments
on the manuscript.

\pagebreak

\noindent
{\bf Figure Captions:}

\noindent
{\bf Fig. 1}:
(a) The CNT-substrate system is divided into three regions {\bf S},
{\bf D} and {\bf L}. (b) The potential profile versus position of a 
double barrier resonant tunneling structure. An electron can be 
transmitted from the continuum states on the top to the bottom, only 
via the localized state (represented by {\bf l}) in the well.
(c) In the absence of defects, the localized and continuum states in
the nanotube are decoupled. In this figure, they are shown spatially
separated for clarity. Coupling between the substrate and cap
causes opening of transport paths where an electron incident in the
substrate tunnels into and out of the localized state before being
scattered into the continuum. This results in an antiresonance.
(d) The presence of defects (represented by {\bf X}) in the tube open
additional transport
paths similar to those in double barrier resonant tunneling structures,
with coupling to the substrate and scattering by the defect acting as
the two scattering centers [compare with (b), where the two scattering
centers are the barriers]. This transforms the transmission 
antiresonance in (c) to a resonance.

\noindent
{\bf Fig. 2}: (10,10) carbon nanotube with a polyhedral cap.
The dashed lines connect equivalent sites of the cap and nanotube in
this two dimensional representation.
The dashed box shows a Stone-Wales defect.

\noindent
{\bf Fig. 3}:
The LDOS at four different cap atoms versus energy for a nanotube 
without defects in contact with the substrate. 
The LDOS is plotted at the cap atom making contact to the substrate,
in the case of a defect-free nanotube  [Fig. 1(c)]. The resonant peaks
correspond to localized energy levels. The four curves correspond to 
the indexed cap atoms in Fig. 2. The legend 'average' corresponds to 
the LDOS averaged over all cap atoms when atom 4 makes contact to the
substrate. $\Sigma_S^r=- 12.5\:i\:$meV ($\Gamma_S=25\:$meV) for the four
cases.

\noindent
{\bf Fig. 4}:
The transmission probability versus energy corresponding to Fig. 3.
The antiresonances occur at the same energy as the LDOS resonances in
Fig. 3. The inset shows an expanded view of the antiresonances.

\noindent
{\bf Fig. 5}:
The width of the antiresonance increases with increase in the strength
of coupling $\Gamma_S$ but the minimum is zero. The dotted and
dashed curves are scaled by 25 and 5 times of the computed transmission 
probability respectively. In these calculations atom 4 makes contact
to the substrate with a coupling strength $\Sigma_S^r=-i\:\Gamma_S/2$,
where the values of $\Gamma_S$ are given in the figure legend.

\noindent
{\bf Fig. 6}:
A one dimensional model describing tunneling between a tip with a
localized energy level $\epsilon_l$ and a substrate. The localized   
level is shown separated from the atom at the edge of the tip for 
clarity.

\noindent
{\bf Fig. 7}:
(a) Transmission probability versus energy when $t_{lc}=0$ shows an
antiresonance. (b) When $t_{lc} \neq 0$, the antiresonance in (a)
disappears around the localized energy level and the transmission
probability has a resonance. The width of this resonance decreases
as $t_{lc}$ decreases.

\noindent
{\bf Fig. 8}: 
(a) Transmission probability versus energy in the absence of a defect.
(b) Transmission probability versus energy for the same structure in
(a) but with a defect located at $L_D$. The strong resonance caused by
an appropriately placed defect is capable of carrying large current. 
In these calculations, the coupling strength between atom 1 and the 
substrate is assumed to be $\Sigma_S^r=- 25\:i\:$meV
($\Gamma_S=50\:$meV).

\begin{figure}[h]
\centerline{\psfig{file=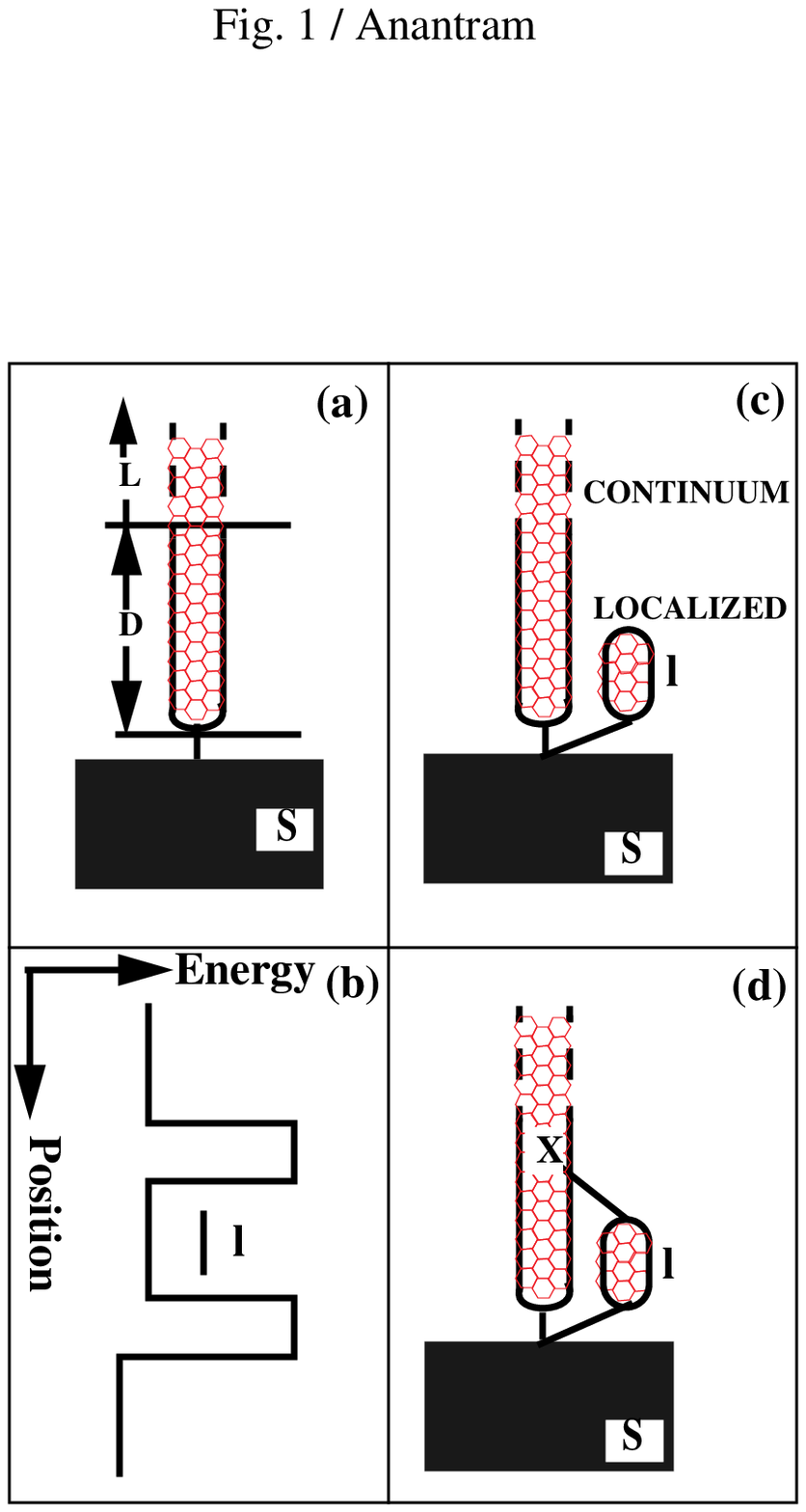,width=5in}}
\small
\end{figure}

\pagebreak

\begin{figure}[h]
\centerline{\psfig{file=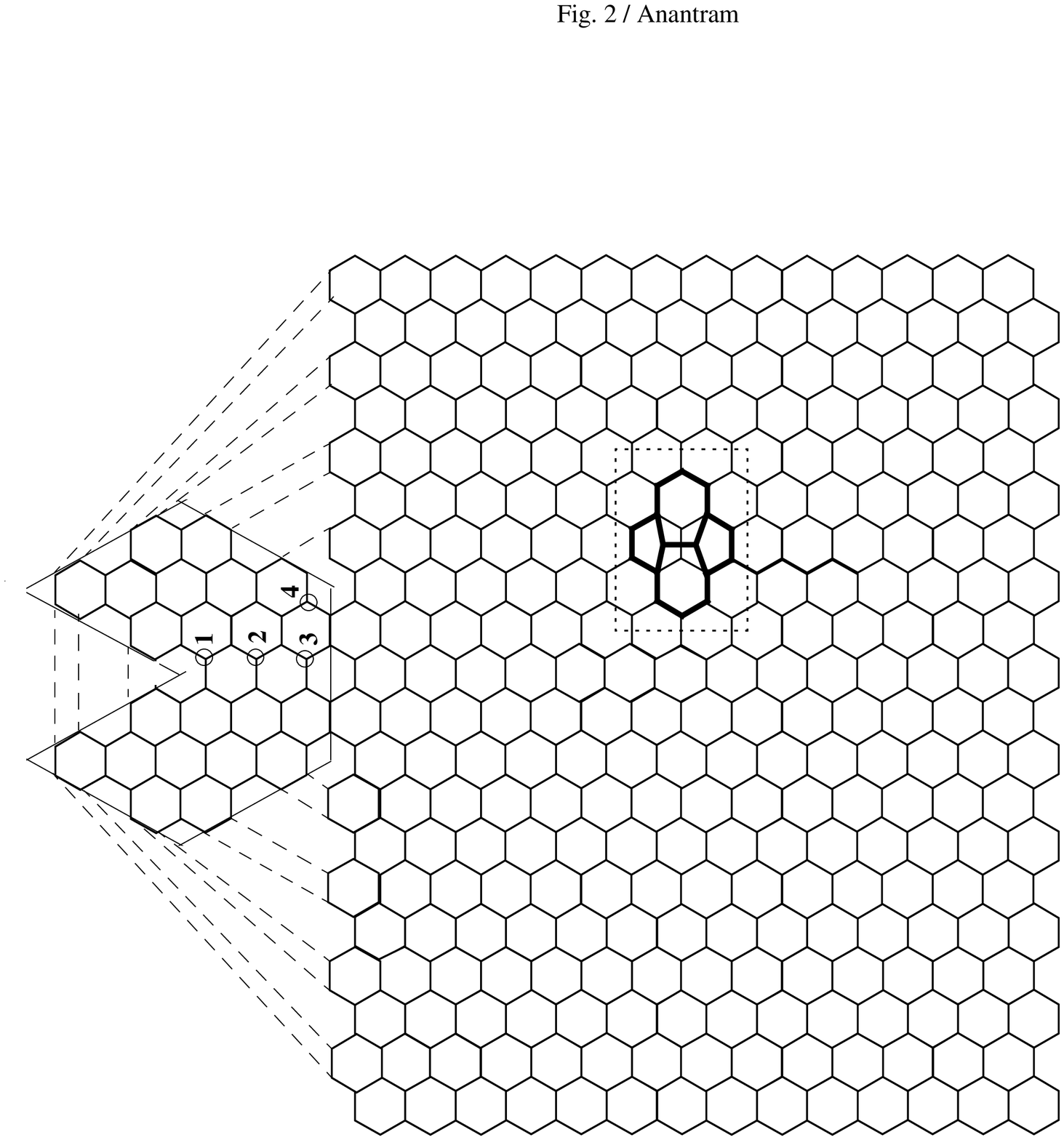,width=7.5in}}
\small
\end{figure}

\begin{figure}[h]
\centerline{\psfig{file=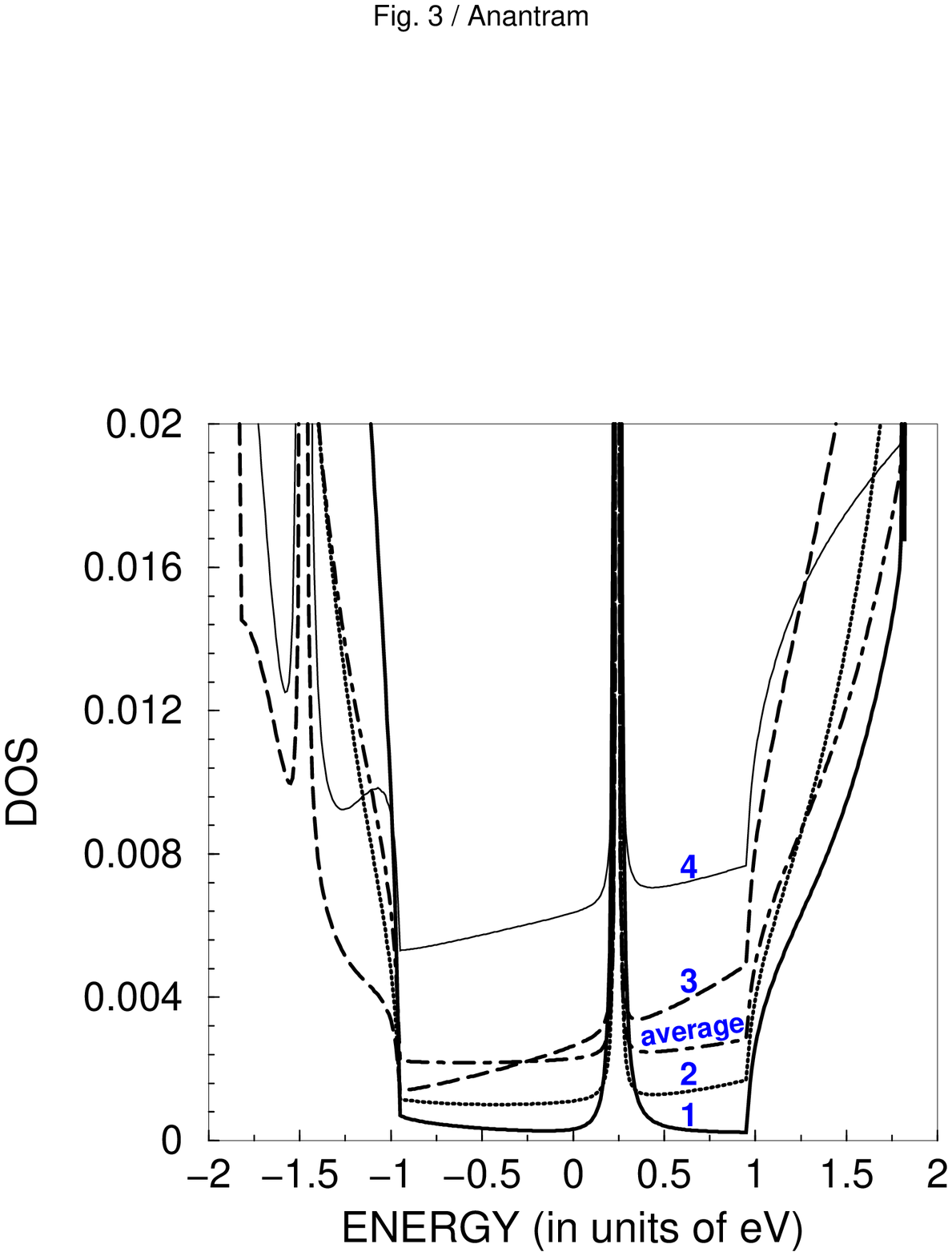,width=4in}}
\small
\end{figure}

\pagebreak

\begin{figure}[h]
\centerline{\psfig{file=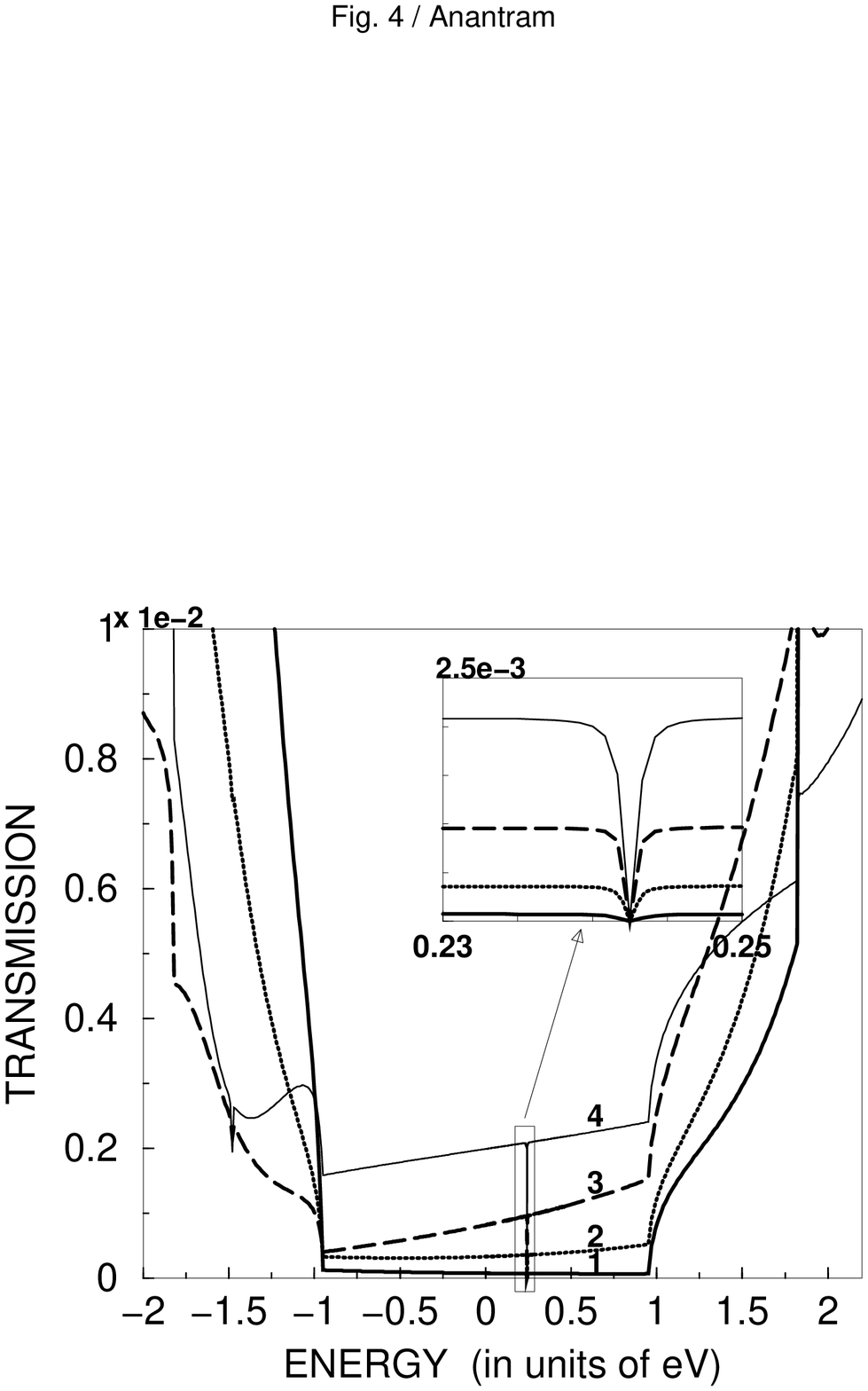,width=4in}}
\small
\end{figure}

\begin{figure}[h]
\centerline{\psfig{file=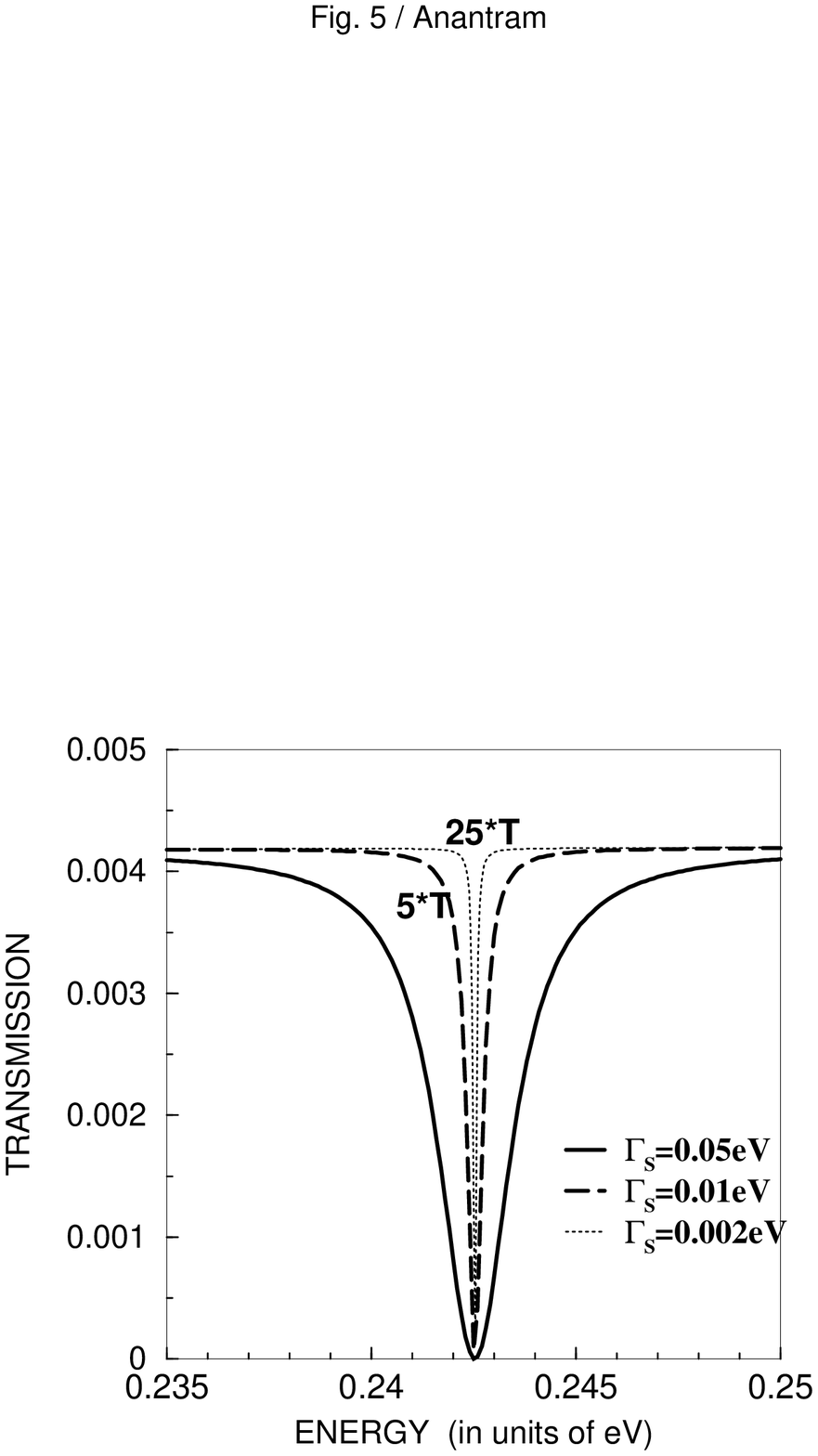,width=4in}}
\small
\end{figure}

\pagebreak

\begin{figure}[h]
\centerline{\psfig{file=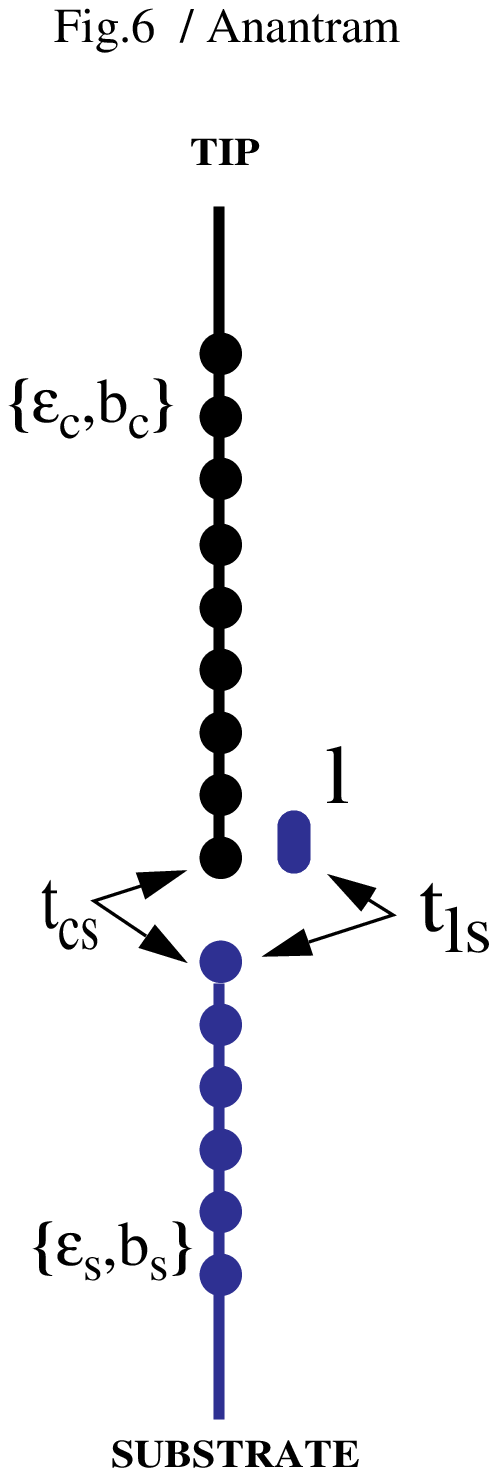,width=4in}}
\small
\end{figure}

\pagebreak

\begin{figure}[h]
\centerline{\psfig{file=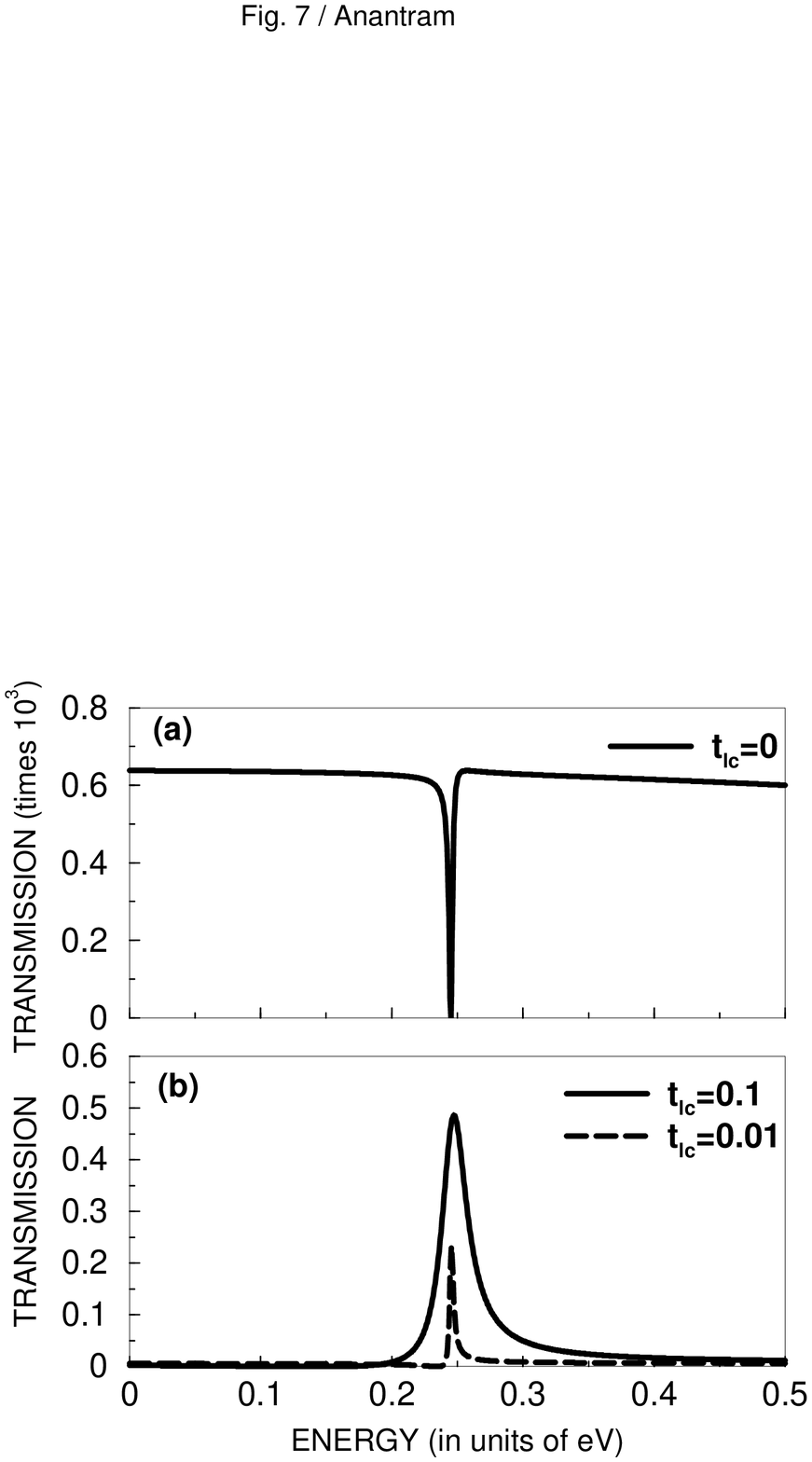,width=4in}}
\small
\end{figure}

\begin{figure}[h]
\centerline{\psfig{file=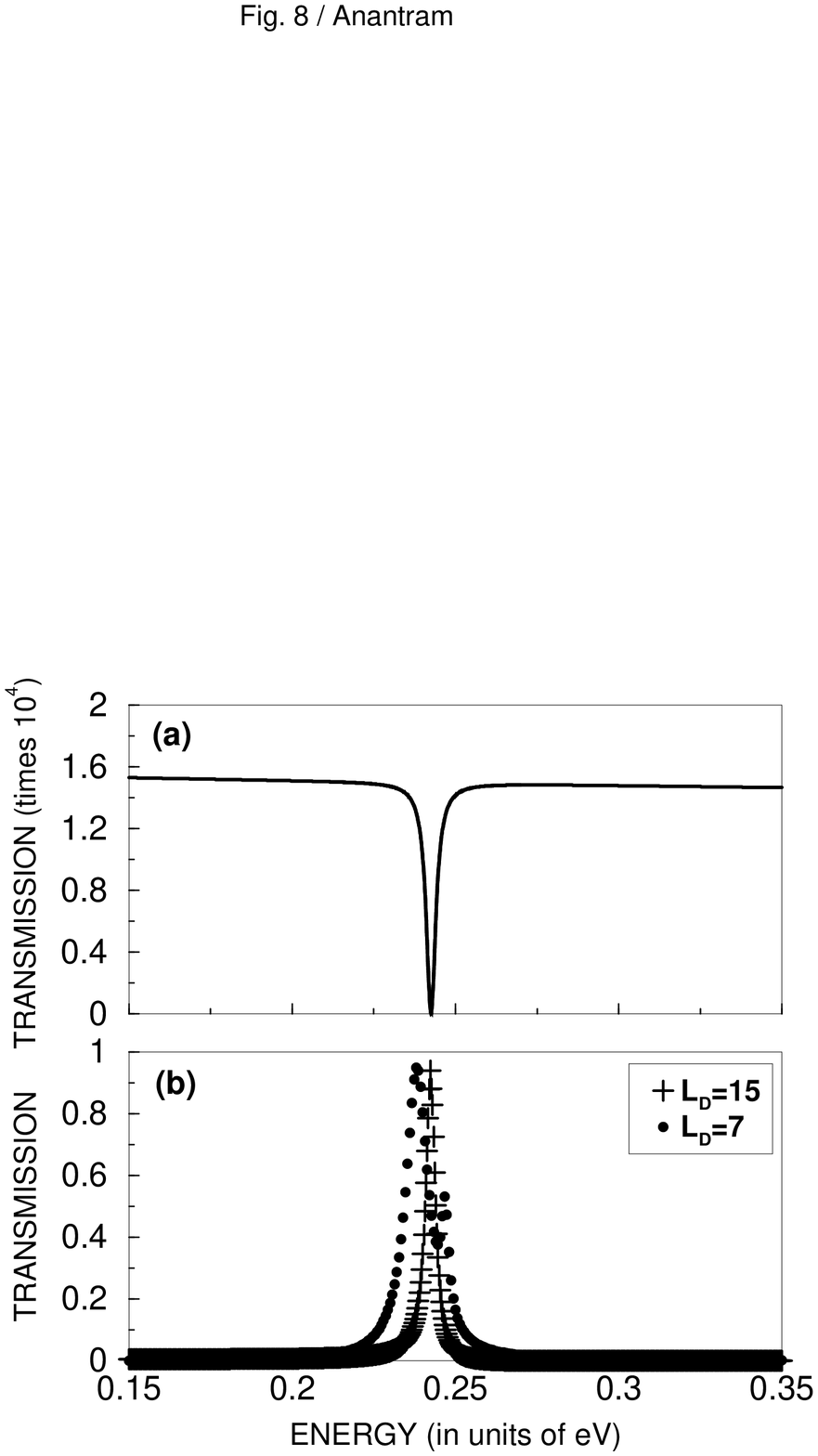}}
\small
\end{figure}


\begin{thebibliography}{100}
\bibitem{comment1}
In references \onlinecite{Dai96,Wong98b,Wong98a},
the carbon nanotube tip is used as an AFM probe rather than an
STM probe.
\bibitem{footnote1}
The interaction between the tip and substrate can also be mediated by
a functional group as demonstrated in Ref. \onlinecite{Wong98a}.
\end{thebibliography}
\end{document}